\documentclass[a4paper,11pt]{article}
\pdfoutput=1 

\usepackage{jcappub} 

\usepackage[T1]{fontenc} 

\subheader{\hfill{\tt IFIC/16-10, TTK-16-23}}

\title{\boldmath Relic neutrino decoupling with flavour oscillations revisited}

\author[a,b]{Pablo F.\ de Salas}
\author[a]{and Sergio Pastor}

\affiliation[a]{Instituto de F\'{\i}sica Corpuscular
(CSIC-Universitat de Val\`{e}ncia)\\ 
Parc Cient\'{\i}fic UV, C/ Catedr\'atico Jos\'e Beltr\'an, 2\\ E-46980 Paterna (Valencia), Spain}

\affiliation[b]{Institute for Theoretical Particle Physics and Cosmology (TTK)\\
RWTH Aachen University, D-52056 Aachen, Germany}

\emailAdd{pabferde@ific.uv.es}
\emailAdd{pastor@ific.uv.es}

\abstract{
We study the decoupling process of neutrinos in the early universe in the presence of 
three-flavour oscillations. The evolution of the neutrino spectra is found by solving the 
corresponding momentum-dependent kinetic equations for the neutrino density matrix, 
including for the first time the proper collision integrals for both diagonal and off-diagonal 
elements. This improved calculation modifies the evolution of the off-diagonal elements of
the neutrino density matrix and changes the deviation from equilibrium of the frozen neutrino 
spectra. However, it does not vary the contribution of neutrinos to the cosmological energy density in the form of radiation, usually expressed in terms of the effective number of neutrinos, $N_{\rm eff}$.
We find a value of $N_{\rm eff}=3.045$, in agreement with previous theoretical calculations and consistent with the latest analysis of Planck data. This result does not depend on the ordering of neutrino masses. We also consider the effect of non-standard neutrino-electron interactions (NSI), predicted in many theoretical models where neutrinos acquire mass. For two sets of NSI parameters allowed by present data, we find that $N_{\rm eff}$ can be reduced down to $3.040$ or enhanced up to 
$3.059$.}
  
\begin{document}
\maketitle
\flushbottom

\section{Introduction}
\label{sec:intro}

A generic prediction of the hot big bang model is the existence of a 
relic background of neutrinos, in number almost as abundant as 
the relic photons that form the cosmic microwave background (CMB).
These neutrinos were produced in the early universe, when its temperature $T$ was 
large enough so that weak interactions were effective and neutrinos were in
thermal contact with charged leptons and the rest of the primeval plasma.
But below the so-called neutrino decoupling temperature, at $T_{\rm dec}\sim 2-3$ MeV,
these elusive particles do not interact any longer and propagate freely.
In the simple but accurate approximation known as {\em instantaneous decoupling},
neutrinos keep the energy distribution of a relativistic fermion,
unchanged but for the effect of redshift of physical momentum, and do 
not share the entropy release from the annihilation of electron-positron pairs into photons
that occurs at temperatures below $T_{\rm dec}$. In such a case, entropy conservation
leads to the well-known ratio of temperatures of relic photons and neutrinos,
$T_\gamma/T_\nu=(11/4)^{1/3}$, and to a contribution of neutrinos to the cosmological energy density of radiation given, in terms of the effective number of neutrinos, by $N_{\rm eff} =3$
\cite{NuCosmo}.

The fact that neutrino decoupling and $e^{\pm}$ annihilations are processes quite close in time
leads to the existence of relic interactions between electrons, positrons and neutrinos
at cosmological temperatures smaller than $T_{\rm dec}$. These processes are more efficient for neutrinos with larger momenta, leading to non-thermal distortions in the neutrino spectra at the percent level and a slightly smaller increase of the comoving photon
temperature, as calculated in \cite{Dicus:1982bz,Herrera1989,Rana91,Dodelson:1992km,Hannestad,dolgov_hansen,Gnedin:1997vn,dolgov_hansen2,esposito,Mangano:2001iu,Mangano:2005cc,Birrell:2014uka,Grohs:2015tfy}. The decoupling process of neutrinos
was studied in these previous analyses with increasing precision, in particular with respect to the numerical solution of the kinetic equations for the neutrino distribution functions, including
finite temperature QED radiative corrections and, in some papers, flavour neutrino oscillations. For instance, in \cite{Mangano:2005cc} it was shown that neutrino mixing is important for fixing the final
flavour spectral distortions and their small effect on primordial nucleosynthesis, but oscillations
do not modify the contribution of neutrinos to the radiation energy density, that was found to be
$N_{\rm eff} =3.046$. Other recent works have obtained, in absence of neutrino oscillations but
including QED corrections, values between $3.044$ \cite{Birrell:2014uka} and $3.052$ \cite{Grohs:2015tfy}. 

The value found for $N_{\rm eff}$ is thus only $0.04-0.05$ larger than the expected number in the instantaneous decoupling approximation, but it fixes the contribution of neutrinos in 
the standard case that must be included in the cosmological model. Only very recently the analysis of cosmological data has achieved a level of precision that is clearly below one
unit in $N_{\rm eff}$. In particular, the data on CMB anisotropies from the Planck satellite, in combination with baryon acoustic oscillations (BAO) measurements, lead to an allowed range of $N_{\rm eff} =3.15\pm 0.23$ (at 68\% CL, data from Planck TT+lowP+BAO), that is restricted to $N_{\rm eff} =3.04\pm 0.18$ if the complete polarization likelihood from Planck is used (see \cite{Ade:2015xua} for all details on the different analyses and combinations of data). Forthcoming cosmological data from future CMB experiments, large-volume galaxy surveys, etc, are expected to improve the sensitivity on $N_{\rm eff}$, possibly down to values such as $0.02-0.04$.

Prompted by the increased accuracy on the measurement of $N_{\rm eff}$, in this paper we revisit the process of neutrino decoupling in the early universe 
with the aim of checking some of the approximations assumed in previous works. In particular, here we solve the Boltzmann equations for the neutrino density matrix
with full collision terms, for both the diagonal and off-diagonal terms, avoiding for the latter the so-called damping factors used in other analyses. We also consider
the present best-fit values of neutrino mixing parameters and investigate whether the results depend on the two possible ways of ordering the neutrino masses, still allowed
by current oscillation experiments. In addition, we take the opportunity to update the analysis performed in  \cite{Mangano:2006ar} and consider how neutrino decoupling 
is affected by the presence of non-standard neutrino-electron interactions.

This paper is organized as follows. In Sec.\ \ref{sec:dec} we present the Boltzmann equations that we need to solve in order to follow the evolution of the neutrino spectra, for both the standard scenario and in the presence of non-standard neutrino-electron interactions. We discuss the approximations with respect to previous works, and comment on the main computation and technical issues.
Our main results on the decoupling of cosmological neutrinos are described in Sec.\ \ref{sec:res},
where we also include the values obtained for the final neutrino spectra and the corresponding contribution to $N_{\rm eff} $ for each case. Finally, we present our conclusions.

\section{Neutrino decoupling including three-flavour oscillations}
\label{sec:dec}

\subsection{Boltzmann equations}
\label{subsec:eq}

The process of neutrino decoupling in the early universe takes place at a temperature of the order of MeV, when weak interactions are no longer effective to keep neutrinos in thermal contact with electrons, positrons, and, indirectly, with photons. It has been shown in previous works (see e.g.\ \cite{Dolgov:2002ab}) that neutrino oscillations become effective at similar temperatures, if the mixing parameters have the values required to explain the data from solar, atmospheric, reactor and accelerator neutrino experiments. In order to take into account the effects of both interactions and oscillations, we consider the evolution of the neutrino density matrix $\varrho_p$,
\begin{equation}\label{densityMatrix}
\varrho_p (t)= \left(\begin{matrix}
\varrho_{ee} & \varrho_{e\mu} & \varrho_{e\tau}\\
\varrho_{\mu e} & \varrho_{\mu\mu} & \varrho_{\mu\tau}\\
\varrho_{\tau e} & \varrho_{\tau\mu} & \varrho_{\tau\tau}
\end{matrix}\right) =
\left(\begin{matrix}
f_{\nu_e} & a_1 + ia_2 & b_1 + i b_2\\
a_1 - i a_2 & f_{\nu_\mu} & c_1 + i c_2\\
b_1 - i b_2 & c_1 - i c_2 & f_{\nu_\tau}
\end{matrix}\right)\, ,
\end{equation}
where the diagonal terms are the occupation numbers $f_{\nu_\alpha}$ of flavour neutrinos with momentum $p$ and the off-diagonal terms, nonzero in the presence of mixing, are described by the real parameters $a_i$, $b_i$ and $c_i$ $(i=1,2)$. Since we neglect a potential neutrino asymmetry, quite constrained due to the measured value of the mixing angle $\theta_{13}$ 
\cite{Mangano12,Castorina:2012md}, neutrinos and antineutrinos share the same density matrices.

The equations of motion for the neutrino density matrices are the corresponding set of Boltzmann equations in an expanding universe \cite{sigl_raffelt} (see also \cite{Blaschke:2016xxt})

\begin{equation}\label{boltzmann}
\left( \partial_t - H p\, \partial_p \right) \varrho_p(t)= -i \left[ \left(\frac{1}{2p} \mathbb{M}_{\rm F} -  \frac{8\sqrt{2} G_{\rm F} p}{3 m^2_{\rm W}} \mathbb{E} \right),\varrho_p (t)\right] + \mathcal{I}\left[ \varrho_p (t)\right]\; ,
\end{equation}
where $H$ is the Hubble parameter, $G_{\rm F}$ is the Fermi constant and $m_{\rm W}$ the W boson mass.
The commutator term includes the vacuum oscillation term proportional to $\mathbb{M}_{\rm F}$, the mass-squared matrix in the flavour basis that is related to the diagonal
one in the mass basis ${\rm diag}(m^2_1,m^2_2,m^2_3)$ via the neutrino mixing matrix $U$, as shown in \cite{Mangano:2005cc}. Here we consider the following values for the neutrino mixing parameters (assuming CP conservation)
\begin{eqnarray}\label{mixpar}
\left(\frac{\Delta m^2_{21}}{10^{-5}~{\rm eV}^2}, \frac{\Delta m^2_{31}}{10^{-3}~{\rm eV}^2},
s^2_{12},s^2_{23},s^2_{13}\right)_{\rm NH} &= &(7.60,2.48,0.323,0.567,0.0226)
\nonumber \\
\left(\frac{\Delta m^2_{21}}{10^{-5}~{\rm eV}^2}, \frac{\Delta m^2_{31}}{10^{-3}~{\rm eV}^2},
s^2_{12},s^2_{23},s^2_{13}\right)_{\rm IH} &= &(7.60,-2.38,0.323,0.573,0.0229)
\end{eqnarray}
where $s^2_{ij}\equiv \sin^2\theta_{ij}$, that correspond to the best-fit values from the global analysis of neutrino oscillation experiments in \cite{Forero:2014bxa} (using the values found in more recent analyses \cite{Gonzalez-Garcia:2015qrr,Capozzi:2016rtj} does not modify our results). The two choices correspond to the two ways of ordering neutrino masses: normal (NH) and inverted (IH) hierarchy.

The second term inside the commutator in eq.~\eqref{boltzmann} corresponds to neutrino forward scattering in the primeval medium, proportional to $G_{\rm F}$ and the diagonal matrix $\mathbb{E}$, that represents the energy densities of charged leptons. For the range of temperatures we are interested in, only the contribution of electrons and positrons is relevant (i.e.\ the only nonzero component of $\mathbb{E}$ is $\mathbb{E}_{11}=\rho_{ee}\equiv\rho_{e^-}+\rho_{e^+}$). We neglect other refractive terms that appear in the general form of eq.~\eqref{boltzmann}: the usual term proportional to the charged-lepton asymmetries (always smaller than either the vacuum or the $\mathbb{E}$ term) and the contribution of neutrino-neutrino interactions, that vanishes for zero neutrino-antineutrino asymmetry.

The last term in eq.~\eqref{boltzmann} includes the effect of non-forward neutrino interactions, proportional to $G_{\rm F}^2$. The most general form of $\mathcal{I}[ \varrho_p (t)]$, for each of the interaction process with neutrinos, is a matrix with a collision integral for each of the components of
$\varrho_p$ \cite{sigl_raffelt}. For instance, the collision term from the annihilation process 
$\nu(p_1)+\bar{\nu}(p_2) \leftrightarrow e^-(p_3)+e^+(p_4)$ is 
\begin{align}
\mathcal{I}_{\nu \bar{\nu} \rightarrow e^- e^+} &= \frac{1}{2}\frac{2^5 G_{\rm F}^2}{2 |\vec{p}_1|} \int \frac{d^3 \vec{p}_2}{(2\pi)^3 2 |\vec{p}_2|}\frac{d^3 \vec{p}_3}{(2\pi)^3 2 E_3}\frac{d^3 \vec{p}_4}{(2\pi)^3 2 E_4} (2\pi)^4 \delta^{(4)} (p_1 + p_2 - p_3 - p_4) \nonumber\\
& \times \left\{ 4(p_1\cdot p_4)(p_2\cdot p_3) F^{LL}_{\rm ann}(\nu^{(1)}, \bar{\nu}^{(2)},e^{(3)}, \bar{e}^{(4)})  \right. \nonumber\\&+ 4 (p_1\cdot p_3) (p_2\cdot p_4) F^{RR}_{\rm ann}(\nu^{(1)}, \bar{\nu}^{(2)},e^{(3)}, \bar{e}^{(4)})  \nonumber\\
& \left. +\, 2 (p_1\cdot p_2) m_e^2 \left( F^{RL}_{\rm ann}(\nu^{(1)}, \bar{\nu}^{(2)},e^{(3)}, \bar{e}^{(4)}) + F^{LR}_{\rm ann}(\nu^{(1)}, \bar{\nu}^{(2)},e^{(3)}, \bar{e}^{(4)}) \right) \right\}, \label{I_ann}
\end{align}
where $m_e$ is the electron mass and we have defined
\begin{align}\label{Fann}
F^{ab}_{\rm ann}(\nu^{(1)}, \bar{\nu}^{(2)},e^{(3)}, \bar{e}^{(4)}) &= f_3 \bar{f}_4 \left( G^a (1-\bar{\varrho}_2) G^b (1-\varrho_1) +(1-\varrho_1) G^b (1-\bar{\varrho}_2) G^a \right) \nonumber\\
&-\, (1-f_3) (1-\bar{f}_4) \left( \varrho_1 G^b \bar{\varrho}_2 G^a + G^a \bar{\varrho}_2 G^b \varrho_1 \right)\;.
\end{align}
Here $f_i=f_{e^-}(p_i)$ and $\bar{f}_i=f_{e^+}(p_i)$ are the distribution functions of charged leptons 
and $G^a$ is a matrix of couplings ($a=L,R$, left-handed $L$ or right-handed $R$). In the absence of non-standard neutrino interactions, $G^{L,R}$ are diagonal with components
\begin{equation}
G^L={\rm diag}\left( g_L, \tilde{g}_L, \tilde{g}_L \right),   \qquad\qquad G^R={\rm diag}\left(g_R, g_R, g_R\right),
\end{equation}
where
\begin{equation}
g_L=\frac{1}{2}+ \sin^2 \theta_{\rm W}, \qquad\qquad \tilde{g}_L = g_L -1, \qquad\qquad g_R=\sin^2 \theta_{\rm W},
\end{equation}
with $\theta_{\rm W}$ the weak mixing angle. 
For the scattering process with electrons we have
\begin{align}
\mathcal{I}_{\nu e^- \rightarrow \nu e^-} &= \frac{1}{2}\frac{2^5 G_{\rm F}^2}{2 |\vec{p}_1|} \int \frac{d^3 \vec{p}_2}{(2\pi)^3 2 E_2}\frac{d^3 \vec{p}_3}{(2\pi)^3 2 |\vec{p}_3|}\frac{d^3 \vec{p}_4}{(2\pi)^3 2 E_4} (2\pi)^4 \delta^{(4)} (p_1 + p_2 - p_3 - p_4) \nonumber\\
& \times \left\{ 4 (p_1\cdot p_4) (p_2 \cdot p_3) F^{RR}_{\rm sc} (\nu^{(1)}, e^{(2)}, \nu^{(3)}, e^{(4)}) \right. \nonumber\\
&+ 4 (p_1 \cdot p_2) ( p_3 \cdot p_4) F^{LL}_{\rm sc} (\nu^{(1)}, e^{(2)}, \nu^{(3)}, e^{(4)})\nonumber\\
& \left. -\, 2(p_1\cdot p_3) m_e^2 \left( F^{RL}_{\rm sc}(\nu^{(1)}, e^{(2)}, \nu^{(3)}, e^{(4)}) + F^{LR}_{\rm sc}(\nu^{(1)}, e^{(2)}, \nu^{(3)}, e^{(4)}) \right)  \right\}, \label{I_scat_el}
\end{align}
and for the scattering with positrons
\begin{align}
\mathcal{I}_{\nu e^+ \rightarrow \nu e^+} &= \frac{1}{2}\frac{2^5 G_{\rm F}^2}{2 |\vec{p}_1|} \int \frac{d^3 \vec{p}_2}{(2\pi)^3 2 E_2}\frac{d^3 \vec{p}_3}{(2\pi)^3 2 |\vec{p}_3|}\frac{d^3 \vec{p}_4}{(2\pi)^3 2 E_4} (2\pi)^4 \delta^{(4)} (p_1 + p_2 - p_3 - p_4) \nonumber\\
& \times \left\{ 4 (p_1\cdot p_4) (p_2 \cdot p_3) F^{LL}_{\rm sc} (\nu^{(1)}, \bar{e}^{(2)}, \nu^{(3)}, \bar{e}^{(4)}) \right. \nonumber\\
&+ 4 (p_1 \cdot p_2) ( p_3 \cdot p_4) F^{RR}_{\rm sc} (\nu^{(1)}, \bar{e}^{(2)}, \nu^{(3)}, \bar{e}^{(4)}) \nonumber\\
& \left. -\, 2(p_1\cdot p_3) m_e^2 \left( F^{RL}_{\rm sc}(\nu^{(1)}, \bar{e}^{(2)}, \nu^{(3)}, \bar{e}^{(4)}) + F^{LR}_{\rm sc}(\nu^{(1)}, \bar{e}^{(2)}, \nu^{(3)}, \bar{e}^{(4)}) \right)  \right\}. \label{I_scat_po}
\end{align}
In both cases we use the definition
\begin{align}\label{Fscat}
F^{ab}_{\rm sc} (\nu^{(1)}, e^{(2)}, \nu^{(3)}, e^{(4)}) &= f_4 (1-f_2) \left( G^a \varrho_3 G^b (1-\varrho_1) + (1-\varrho_1) G^b \varrho_3 G^a \right)\nonumber\\
&- f_2 (1-f_4) \left( \varrho_1 G^b (1-\varrho_3) G^a + G^a (1-\varrho_3) G^b \varrho_1 \right),
\end{align}
where $f_i=f(p_i)$ is the distribution function of the electron or positron, depending on which particle we are considering for the scattering. The collision terms for the neutrino-neutrino processes are similar to eqs.\ (\ref{I_ann}), (\ref{I_scat_el}) and (\ref{I_scat_po}), but with more complicated expressions that are non-linear in the neutrino density matrices \cite{sigl_raffelt}.


As in previous analyses of neutrino decoupling in the early universe 
\cite{Hannestad,dolgov_hansen,esposito,Mangano:2001iu,Mangano:2005cc,Birrell:2014uka,Grohs:2015tfy}, in our calculations the collision terms for the diagonal components $\varrho_{\alpha\alpha}$ (the distribution functions of flavour neutrinos $f_{\nu_\alpha}$) are solved numerically after they are analytically reduced to two-dimensional integrals, following the process described in \cite{dolgov_hansen}. 
However, previous works have approximated the off-diagonal collision terms 
of $\mathcal{I}_{\alpha\beta}[ \varrho_p]$ in eq.\ \eqref{boltzmann}
as damping factors with an expression $-D_{\alpha\beta}(p,t)
\varrho_{\alpha\beta}$, where $\alpha\neq \beta$ and the $D$ functions 
can be obtained under some approximations in a similar way as in \cite{Enqvist:1991qj,McKellar:1992ja,Bell:1998ds}. With this prescription, one assumes that each time neutrinos
interact in a process that distinguishes among flavors they collapse into weak-interaction eigenstates.
In our analysis we relax that approximation and deal with the off-diagonal collision terms in the same way as for the diagonal variables for the processes that involve neutrinos and $e^\pm$. We do not consider the full off-diagonal terms for weak reactions involving only neutrinos, such as $\nu\nu\leftrightarrow \nu\nu$ or $\nu\bar{\nu}\leftrightarrow \nu\bar{\nu}$, that play a less important role in the process of neutrino heating (see e.g.\ the detailed discussion in \cite{Grohs:2015tfy}). This assumption also allows us to reduce the computing time.

The Boltzmann equations for the neutrino density matrices must be solved simultaneously with
the continuity equation for the total energy density of radiation $\rho$,
\begin{equation}\label{EnergyCons}
\frac{d\rho}{dt} = -3H (\rho + P)\; ,
\end{equation}
where $\rho$ and $P$ are the total energy density and pressure, respectively, of the relativistic plasma, that includes the electromagnetic components $\gamma$ and $e^\pm$ (in equilibrium with temperature $T_\gamma$) and the three neutrino states. This equation gives the time evolution of the photon temperature $T_\gamma$.

Finally, for the cosmological period of interest, finite temperature QED corrections induce effective electron and photon masses, that in turn modify the equation of state of the electromagnetic plasma and lead to small changes (via $\delta m_e^2$) in the collision rates of the processes involving $e^\pm$. These modifications are included in our calculations as described in \cite{heckler,fornengo,Mangano:2001iu}.

\subsection{Non-standard neutrino-electron interactions}

Neutrino decoupling in the early universe would be modified in the presence of interactions beyond the weak processes present in the Standard Model (SM) of particle physics. Since neutrinos are massless in the framework of the SM, many extended models have been proposed were neutrinos acquire mass.
Most of these models naturally lead to new non-standard interactions (NSI) involving neutrinos, whose value strongly depends on the particular model.

In this paper we follow refs.\ \cite{Berezhiani:2001rs,Davidson:2003ha,Mangano:2006ar} and assume that new physics induces NSI only through the four-fermion operators $(\bar{\nu}{\nu})( \bar{f} f)$, where $f$ is a charged lepton or a quark. Moreover, for the decoupling of cosmological neutrinos only the NSI that involve electrons are important. They are described, together with the standard weak interactions, by the effective Lagrangian
\begin{equation}
{\cal L} = {\cal L}_{\rm SM}+\sum_{\alpha,\beta} {\cal L}_{\rm NSI}^{\alpha\beta}
\label{Ltot}
\end{equation}
which, in addition to the SM term \cite{Mangano:2006ar}, contains the contribution from NSI
\begin{equation}
{\cal L}_{\rm NSI}^{\alpha\beta}=-2\sqrt{2}{G_{\rm F}}
\sum_P\varepsilon_{\alpha\beta}^P(\bar{\nu}_\alpha\gamma^\mu L\nu_\beta)(\bar{e}\gamma_\mu Pe)
\label{Lnsi}
\end{equation}
where $\alpha,\beta=e,\mu,\tau$ and $P=L,R=(1\mp \gamma_5)/2$ are the chiral operators. The NSI parameters $\varepsilon_{\alpha\beta}^P$ can lead to a flavour-changing contribution 
($\alpha\neq\beta$) or induce a breaking of lepton universality ($\alpha=\beta$). Their values are constrained by data from various laboratory experiments, as recently reviewed in 
\cite{Miranda:2015dra}, but due to possible cancellations the limits depend on the number of NSI parameters that are simultaneously included.

The effect of NSI on relic neutrino decoupling was studied in \cite{Mangano:2006ar}, where
it was analytically shown how NSI parameters lead, in general, to a larger interaction between neutrinos and $e^\pm$, although in some particular regions of $\varepsilon$ they could reduce the collision rate
(see Fig.\ 3 in \cite{Mangano:2006ar}). In any case, the presence of NSI modifies the evolution
 of the relevant parameters, and in particular the final value of $N_{\rm eff}$. For a particular combination of $\varepsilon$ values, then allowed by experiments, it was shown that the
deviation of $N_{\rm eff}$ from $3$ was three times the enhancement from neutrino heating with SM interactions. 

The Boltzmann equations in eq.~\eqref{boltzmann} are modified due to NSI interactions as follows
\begin{equation}\label{boltzmannNSI}
\left( \partial_t - H p \partial_p \right) \varrho_p(t)= -i \left[ \left(\frac{1}{2p} \mathbb{M}_{\rm F} -  \frac{8\sqrt{2} G_{\rm F} p}{3 m^2_{\rm W}}~\rho_{ee} \mathbb{E}_{\rm NSI} \right),\varrho_p (t)\right] + \mathcal{I}\left[ \varrho_p (t), \varepsilon\right]\; ,
\end{equation}
where the matrix  
\begin{equation}
 \mathbb{E}_{\rm NSI}=\left(\begin{matrix}
1 + \varepsilon_{ee} & \varepsilon_{e\mu} & \varepsilon_{e\tau}\\
\varepsilon_{e\mu} & \varepsilon_{\mu\mu} & \varepsilon_{\mu\tau}\\
\varepsilon_{e\tau} & \varepsilon_{\mu\tau} & \varepsilon_{\tau \tau}
\end{matrix}\right)\; ,
\end{equation}
is no longer diagonal and contains the effect of non-standard interactions via the combinations of NSI parameters 
\begin{equation}
\varepsilon_{\alpha\beta} = \varepsilon_{\alpha\beta}^L + \varepsilon_{\alpha\beta}^R.
\end{equation}
The presence of NSI also modifies the statistical factors of eqs.~\eqref{Fann} and \eqref{Fscat}
in the collision integrals. The coupling matrices adopt the form
\begin{equation}
G^L = \left(\begin{matrix}
g_L + \varepsilon_{ee}^L & \varepsilon_{e\mu}^L & \varepsilon_{e\tau}^L\\
\varepsilon_{e\mu}^L & \tilde{g}_L + \varepsilon_{\mu\mu}^L & \varepsilon_{\mu\tau}^L \\
\varepsilon_{e\tau}^L & \varepsilon_{\mu\tau}^L & \tilde{g}_L + \varepsilon_{\tau\tau}^L
\end{matrix}\right), \qquad\qquad G^R = \left(\begin{matrix}
g_R + \varepsilon_{ee}^R & \varepsilon_{e\mu}^R & \varepsilon_{e\tau}^R\\
\varepsilon_{e\mu}^R & g_R +\varepsilon_{\mu\mu}^R & \varepsilon_{\mu\tau}^R \\
\varepsilon_{e\tau}^R & \varepsilon_{\mu\tau}^R & g_R + \varepsilon_{\tau\tau}^R
\end{matrix}\right).
\end{equation}

In our analysis we will restrict the calculation of neutrino decoupling with NSI and full collision terms, as described in sec.\ \ref{subsec:eq}, to two combinations of $\varepsilon$ parameters with $\varepsilon_{ee}^P$ and 
$\varepsilon_{\tau\tau}^P$, that are still allowed by present data \cite{Forero:2011zz,Miranda:2015dra} . We do not consider the contribution of $\varepsilon$ that affect muon-neutrino interactions because they are strongly suppressed \cite{Barranco:2007ej}. 

\subsection{Computation and technical issues}
\label{subsec:comp}

We solve the system of Boltzmann equations \eqref{boltzmann} and the continuity equation \eqref{EnergyCons} with the same comoving variables as in \cite{Mangano:2005cc},
\begin{equation}\label{comoving}
x = m_e\; a \;,\qquad\qquad  y = p\; a \;,\qquad\qquad z= T_\gamma\; a\, ,
\end{equation}
where we have chosen as arbitrary mass scale the electron mass, $p$ is the neutrino momentum, $T_\gamma$ is the photon temperature and $a$ is the cosmological scale factor, normalized so that $a(t) \rightarrow 1/T_\gamma$ at large temperatures. The expressions of the Boltzmann equations in terms of these comoving variables are listed in appendix \ref{Pfunc}.

The system of kinetic equations to be solved is integro-differential due to the presence of the neutrino density matrices in the collision terms. In previous analyses of neutrino decoupling, this system was solved either using a discretization in a grid for the neutrino momenta $y_i$ as in 
\cite{Hannestad,dolgov_hansen,Gnedin:1997vn,dolgov_hansen2,Mangano:2005cc,Grohs:2015tfy}
or expanding the non-thermal distortions of the neutrino distribution functions in moments as in \cite{esposito,Mangano:2001iu,Birrell:2014uka}. Here we choose the first method and use a grid of  values for the neutrino momenta $y_i$ in the range $[0.01,20]$. We found 
good convergence for $100$ bins. The system of integro-differential equations is solved using the free FORTRAN77 library ODEPACK, which includes several solvers for differential equations.

We start the numerical computation at a value $x_{\rm in}=m_e/T_\gamma^0$, where 
$T_\gamma^0=10$ MeV, while neutrinos are still in good thermal contact with the electromagnetic plasma via weak interactions and flavour oscillations are suppressed by the medium. Therefore, the initial values of the off-diagonal components of $\varrho(y_i)$ are zero, while the diagonal terms are 
$[\exp(y_i/z_{\rm in})+1]^{-1}$, with $z_{\rm in}=1.00003$ the initial value of the dimensionless photon temperature. This number is found solving the continuity equation (\ref{EnergyCons}) with neutrinos fully coupled to electrons and positrons \cite{dolgov_hansen2}. The system is solved up to a final value $x_{\rm fin}=30$ when the neutrino distribution functions and $z$ have reached their asymptotic values.

\section{Results}
\label{sec:res}
\begin{figure}[tbp]
\centering 
\includegraphics[width=.99\textwidth]{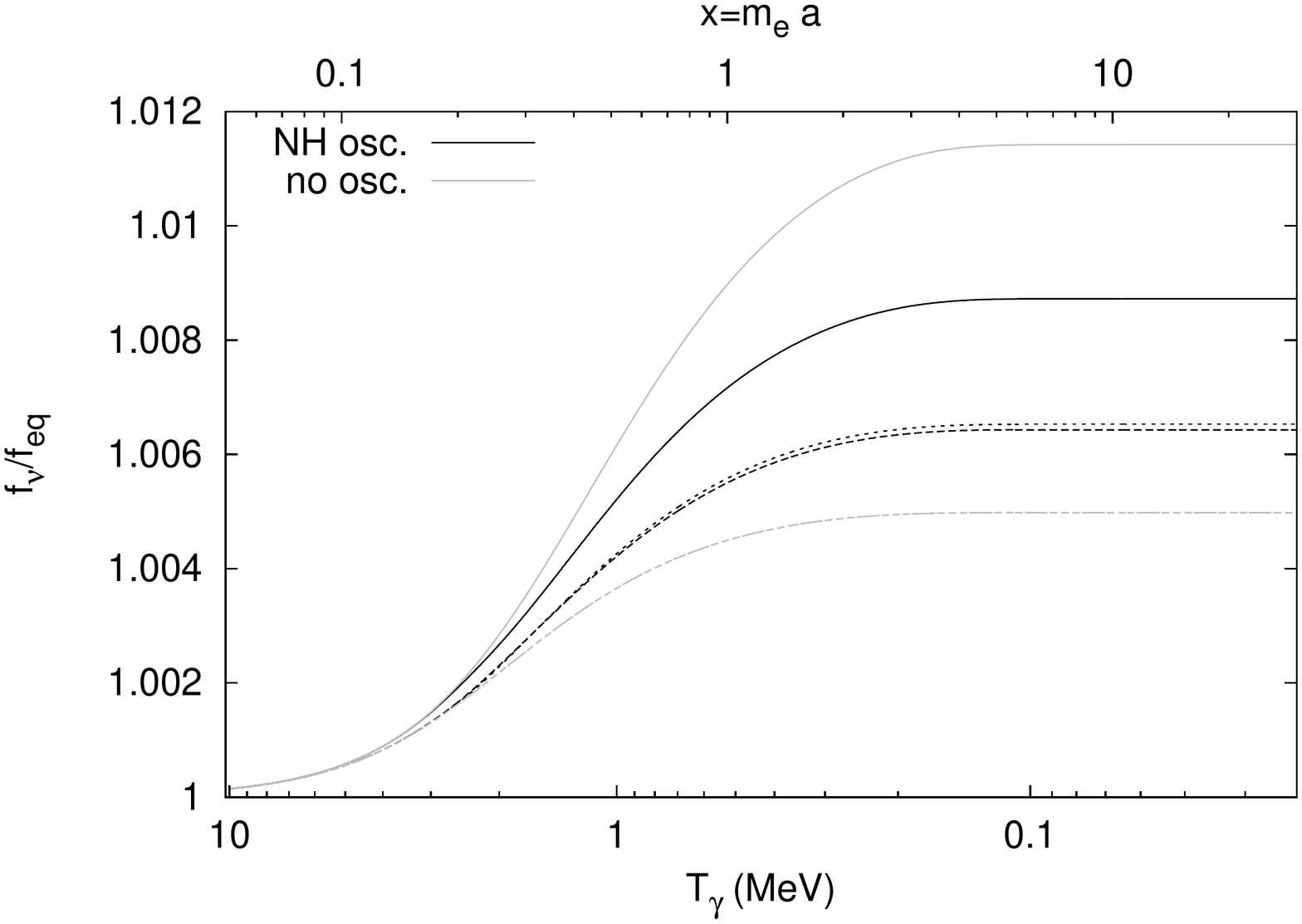}
\caption{\label{fig:evol_i} Evolution of the distortions of the neutrino spectra for the dimensionless momentum ${y=5}$ with standard neutrino interactions, as a function of $x$ or the photon temperature.
Outer (inner) lines correspond to the case with no neutrino mixing (with oscillations and
masses in the NH case). The upper two lines correspond to electron neutrinos and the lower lines to muon and tau neutrinos (slightly different in the case with oscillations).}  
\end{figure}

\subsection{Standard case with flavour oscillations}
\label{subsec:result-standard}

First, we consider the case with standard weak interactions, with and without neutrino oscillations. In the presence of neutrino mixing, we are mainly interested in the possible effects of including the full expressions for the off-diagonal collision terms in the kinetic equations. On the other hand, once the collision terms are fixed, we also want to check whether there are differences between the two options for the neutrino mass hierarchy or the present best-fit values of the mixing parameters lead to modifications with respect to the results in \cite{Mangano:2005cc}.

We show in Fig.\ \ref{fig:evol_i}  the evolution of the flavour neutrino spectra for a particular value of the neutrino momentum ($y=5$). The corresponding distortions $f_{\nu_\alpha}/f_{\rm eq}$, where
$f_{\rm eq}=[\exp(y)+1]^{-1}$, are shown as a function of the photon temperature or the cosmological expansion ($x$). The behaviour of this evolution has been discussed in previous works (see e.g.\ \cite{dolgov_hansen,Mangano:2005cc}). At large temperatures ($T_{\gamma}\gtrsim 2$ MeV) neutrinos are still interacting with $e^\pm$ and 
their energy spectra keeps an equilibrium form with $T_\gamma$. Later the cosmological expansion renders less efficient the weak processes and neutrinos decouple from the electromagnetic plasma in a momentum-dependent way. The residual interactions lead to spectral distortions for neutrinos, which are larger for the electronic flavour. Neutrino oscillations, effective after $T_\gamma \lesssim 3$ MeV when the medium potential is diluted by the expansion, reduce the difference in the spectral distortions of electron neutrinos with respect to the other flavours.
\begin{figure}
\centering 
\includegraphics[width=.99\textwidth]{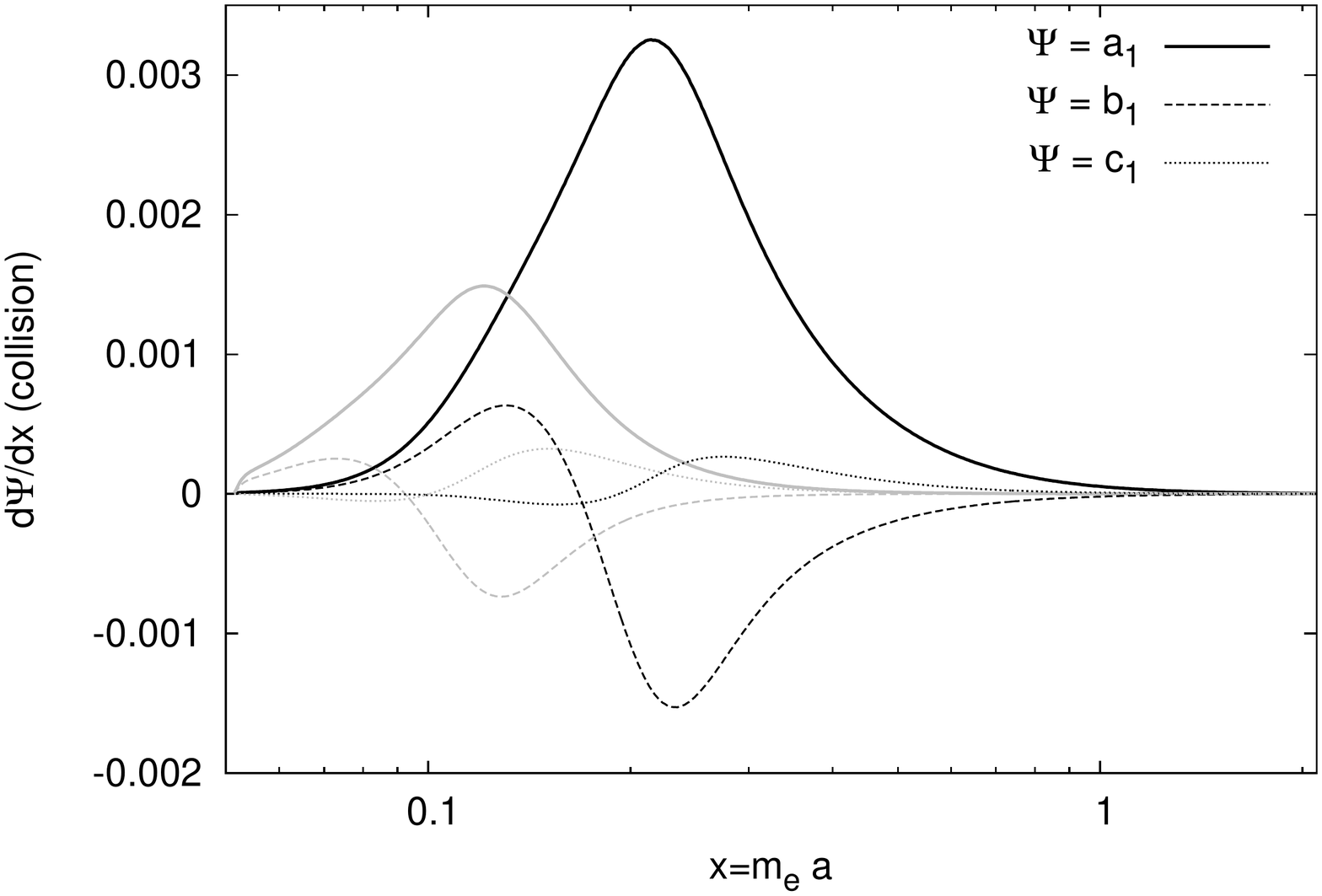}
\caption{\label{fig:evol_off} Collision terms for the real parts  of the off-diagonal elements of the neutrino density matrix in eq.~\eqref{densityMatrix}: $a_1$ (full lines), $b_1$ (dashed lines), and $c_1$ (dotted lines). They are shown for a dimensionless momentum ${y=5}$ and were calculated with the full expression (black lines) or with damping functions (grey lines), accounting only for the processes involving the interactions of neutrinos and $e^\pm$.}
\end{figure}

The results shown in Fig.\ \ref{fig:evol_i} when oscillations are taken into account were found with the full collision terms (the matrix $\mathcal{I}[ \varrho_y]$) for neutrino-electron processes, avoiding the damping functions for the off-diagonal elements. 
In order to have an idea of the difference, we show in 
Fig.\ \ref{fig:evol_off} the evolution of the collision terms for the real parts of the off-diagonal components $\varrho_{\alpha\beta}$ in eq.\ \eqref{densityMatrix}, for a particular neutrino momentum
($y=5$). For each case ($a_1,b_1,c_1$), the result of the full integral is compared with the corresponding damping term. One can see that, while the overall behaviour is similar for each case, the full calculation presents a smoother evolution starting from zero at large temperatures. It is also interesting that the largest (absolute) values of the collision terms with the full integrals appear later than in the simple damping prescription. Since collisional damping leads to 
a loss of flavour coherence, a difference in the off-diagonal terms can modify the 
evolution of neutrino oscillations and eventually change the final distortions in the neutrino spectra.

\begin{figure}
\centering 
\includegraphics[width=.99\textwidth]{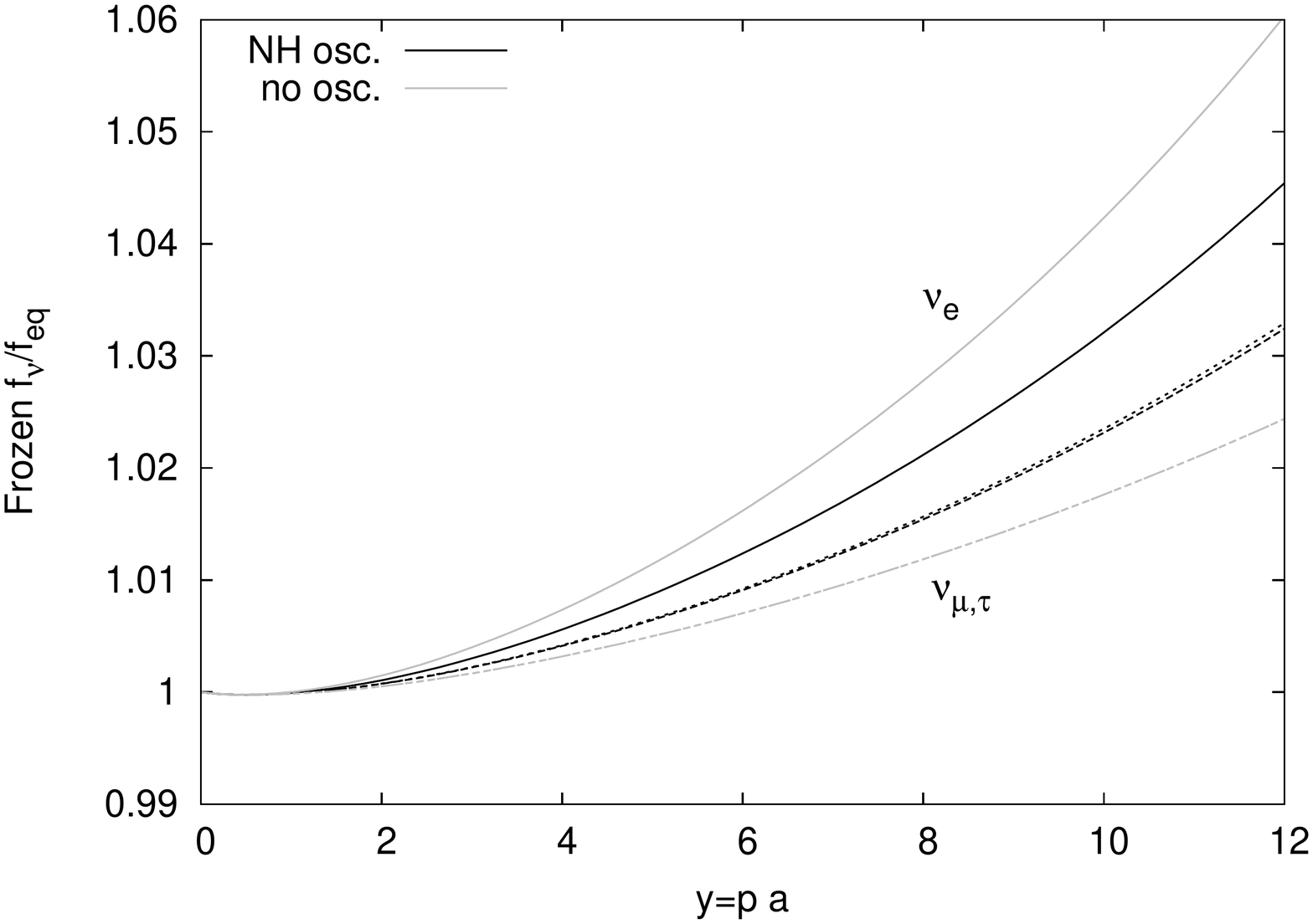}
\caption{\label{fig:delta_alpha} Final distortions of the flavour neutrino spectra as a function of the comoving momentum. Outer (inner) lines correspond to the case with no neutrino mixing (with oscillations and
masses in the NH case). The IH case overlaps the NH one at this scale.}  
\end{figure}

\begin{figure}[tbp]
\centering 
\includegraphics[width=.99\textwidth]{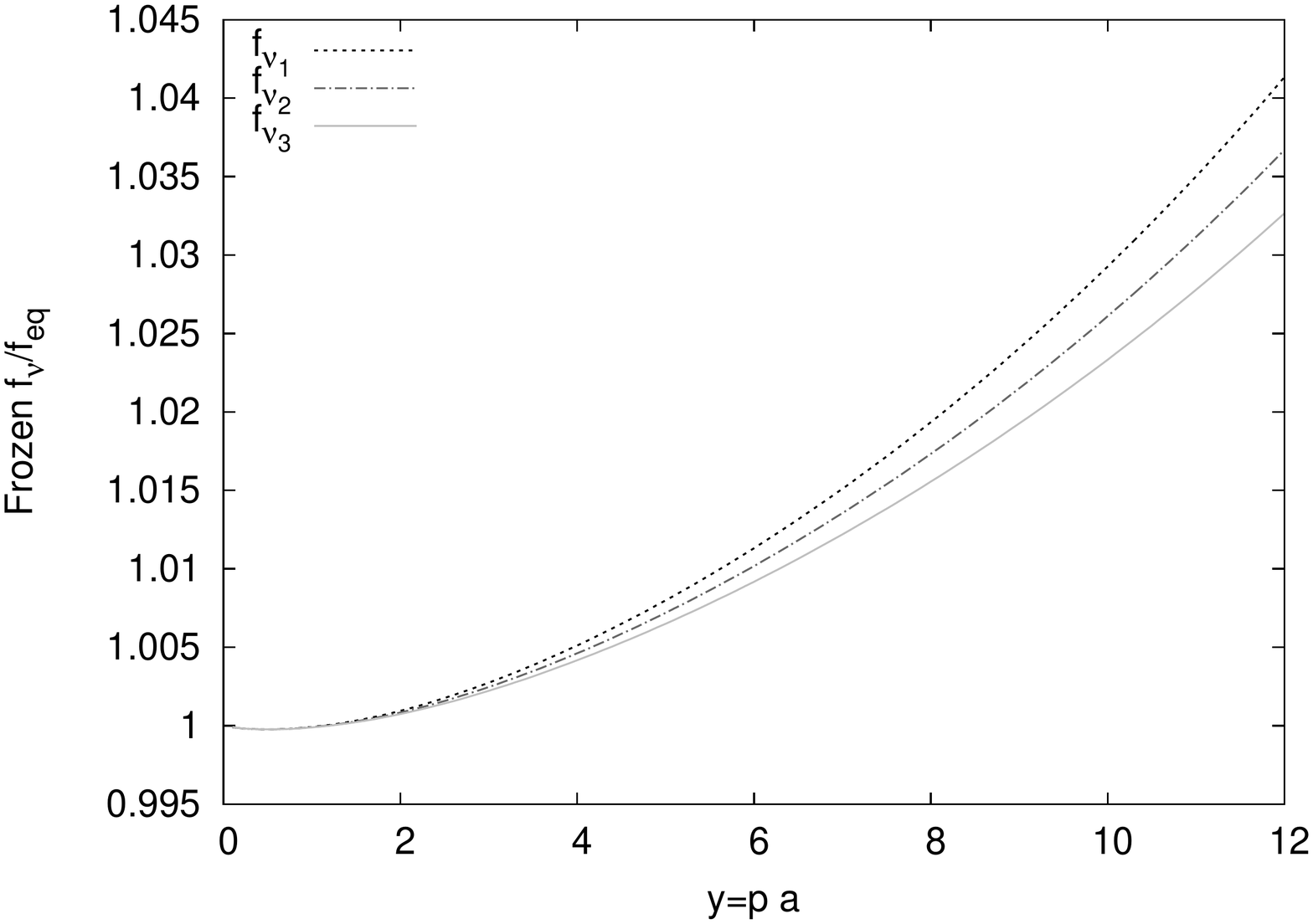}
\caption{\label{fig:delta_i} Same as Figure \ref{fig:delta_alpha} for the mass neutrino states and NH. The spectral distortions of the mass states for the IH case overlap with those for NH at this scale.}
\end{figure}

The final values of the momentum-dependent distortions in the spectra of flavour neutrinos are shown in Fig.\ \ref{fig:delta_alpha}. Their non-thermal character is evident and can be understood from the fact that more energetic neutrinos are interacting with $e^\pm$ at smaller temperatures. The presence of neutrino oscillations leads to a reduced difference between the 
distortions of electron neutrinos and those of the other flavours. The case with oscillations
shown in Fig.\ \ref{fig:delta_alpha} corresponds to the normal neutrino mass hierarchy, but the
calculation with an inverted hierarchy leads to very similar results and the final distortions basically 
overlap with those found with NH.

\begin{table}[tbp]
\centering
\begin{tabular}{|l|ccccc|}
\hline
Case & $z_{\rm fin}$ & $\delta\bar{\rho}_{\nu_e}$(\%) & $\delta\bar{\rho}_{\nu_\mu}$(\%) & 
$\delta\bar{\rho}_{\nu_\tau}$(\%) & $N_{\rm eff}$ \\
\hline
Instantaneous decoupling & $1.40102$ & $0$ & $0$ & $0$ & $3.000$\\
Inst.\ dec.\ (with QED corrections) & $1.39975$ &$0$ & $0$ & $0$ & $3.011$\\
No oscillations & $1.39784$ & $0.920$ & $0.392$ & $0.392$ & $3.045$\\
\hline
Normal hierarchy (NH)     & $1.39779$ & $0.699$ &  $0.511$ & $0.519$ & $3.045$\\
NH (damping terms)    & $1.39778$ & $0.673$ & $0.529$ & $0.533$ & $3.045$\\
Inverted hierarchy (IH)      & $1.39779$ & $0.697$ & $0.512$ & $0.520$ & $3.045$\\
\hline
NSI $ee$  (NH)	& $1.39812$ & $0.592$ & $0.460$ & $0.467$ & $3.040$\\
NSI $\tau\tau$ (NH)  & $1.39708$ & $0.862$ & $0.784$ & $0.812$ & $3.059$\\
\hline
\end{tabular}
\caption{\label{table_alpha} Final values of the dimensionless photon temperature, the distortion of the energy densities of flavour neutrinos and the effective number of neutrinos, computed for several cases as discussed in the text.}
\end{table}

We use the frozen values of the neutrino spectra to calculate how the energy density of each 
flavour state is enhanced with respect to the instantaneous decoupling approximation. The results are given in Table \ref{table_alpha} in terms of $\delta\bar{\rho}_{\nu_\alpha}\equiv 
(\rho_{\nu_\alpha}-\rho_{\nu_0})/\rho_{\nu_0}$, where $\rho_{\nu_0}$ is the energy density of 
one neutrino state that does not share the entropy release from the $e^\pm$ annihilations. In all cases the enhancement is at the sub-percent level, which indicates the well-known result that the non-thermal distortions for cosmological neutrinos are very small. The main effect is a slightly larger contribution of neutrinos to
the cosmological energy density of radiation $\rho_{\rm r}$, usually given, as in the case of any other contribution of relativistic particles other than photons, in terms of the effective number of neutrinos,
\begin{equation}\label{neff}
N_{\rm eff} = \frac{8}{7}\left(\frac{11}{4}\right)^{4/3}
\left[\frac{\rho_{\rm r}-\rho_\gamma}{\rho_\gamma}\right],
\end{equation}
that can be also written as
\begin{equation}\label{neff2}
N_{\rm eff} = \left(\frac{z_0}{z_{\rm fin}}\right)^4
\left(3+\delta\bar{\rho}_{\nu_e}+\delta\bar{\rho}_{\nu_\mu}+\delta\bar\rho_{\nu_\tau}\right)\;.
\end{equation}
Here $z_0=(11/4)^{1/3}$ and $z_{\rm fin}$ are the asymptotic values of the dimensionless photon temperature in the instantaneous decoupling approximation and for each case in our calculations, respectively. The final values of $z_{\rm fin}$ for each case and the calculated ones for $N_{\rm eff}$ are given in Table \ref{table_alpha}. 

Our main results correspond to the cases NH and IH in Table \ref{table_alpha}, computed with the full collision terms. Both cases lead to almost equal values of the enhanced neutrino energy densities and $z_{\rm fin}$, and to the same final effective number of neutrinos,
$N_{\rm eff}=3.045$, which is almost the same as $3.046$, found in \cite{Mangano:2005cc} and usually taken as reference for the standard value of $N_{\rm eff}$. Such a difference ($0.001$) is
smaller than the accuracy quoted in \cite{Mangano:2005cc} and corresponds to what we estimate for our own numerical calculations (performed with a completely independent code).
We conclude that the value of $N_{\rm eff}$ does not depend on the ordering of neutrino masses.
However, the NH and IH cases are not exactly equivalent; there is a resonance for IH when neutrino oscillations become effective ($T_\gamma \sim 3\;{\rm MeV}$) that leads to an interchange among the mass states, although they evolve in a similar way after the resonance for both mass hierarchies. The difference is therefore not appreciable for the range of temperatures after $e^\pm$ have annihilated and neutrinos are completely decoupled, as can be seen in Table \ref{table_i} and Fig. \ref{fig:delta_i}, where our main results are shown for the mass neutrino eigenstates $\nu_{1,2,3}$. The corresponding neutrino distribution functions are the relevant
energy spectra for all cosmological calculations at temperatures much smaller than the neutrino decoupling value (MeV), both when neutrinos are still relativistic or when they start to behave as non-relativistic particles. 

It is interesting to check if, fixing the mass ordering to NH, the differences shown in Fig.\ \ref{fig:evol_off} between the off-diagonal collision terms calculated with the full integrals or the simple damping  prescription, have any effect on the final values that characterize relic neutrino decoupling. We find that the value of $N_{\rm eff}$ is, within the accuracy of our numerical calculations, the same in both cases ($3.045$), while there exist small differences in the values of the energy density distortions $\delta\bar{\rho}_{\nu_\alpha}$, that are slightly smaller (larger) for  muon or tau (electron) neutrinos when the damping approximation is not used. However, one can conclude that for practical purposes, in particular concerning the final value of $N_{\rm eff}$, the inclusion of the full collision integrals in the evolution equation of the off-diagonal components of 
$\varrho_p$ is not necessary.

We have also included in Table \ref{table_alpha} our results for the case when neutrino oscillations are not considered. As shown in previous analyses, the energy densities of the neutrino flavour states in absence of mixing are quite different with respect to those found with neutrino oscillations, but the final value of $N_{\rm eff}$ is again $3.045$. This can be compared with the values found in previous works for zero neutrino mixing but including QED corrections: $3.046$ \cite{Mangano:2005cc}, $3.044$ \cite{Birrell:2014uka} and $3.052$ \cite{Grohs:2015tfy}. While some of the small differences among these values could be due to different methods for the numerical evaluation, we believe that a new, specific study of the effects of finite temperature QED radiative corrections on relic neutrino decoupling could be useful in order to understand the approximations assumed.

\subsection{Non-standard neutrino-electron interactions}

Here our aim is not to explore the whole space of NSI parameters allowed by laboratory data, but to show the effects of the new interactions on relic neutrino decoupling with a couple of examples.
We consider two sets of diagonal parameters $\varepsilon_{ee}^P$ and $\varepsilon_{\tau\tau}^P$ that are not ruled out by experimental data, as described in \cite{Forero:2011zz}.
The two chosen sets are $\varepsilon_{\tau\tau}^R = -\varepsilon_{\tau\tau}^L = 0.37$; and $\varepsilon_{ee}^R = -0.42$ and $\varepsilon_{ee}^L = -0.09$, setting in both cases the rest of NSI parameters to zero, and including neutrino oscillations with a normal ordering of neutrino masses.


We chose these values because they correspond to allowed NSI parameters that are the furthest from the standard case, where all $\varepsilon=0$. In addition, with this selection of values we want to illustrate the two possible effects on neutrino decoupling: the first case ($\varepsilon_{\tau\tau}\neq 0$) leads to an increase in the effective number of neutrinos with respect to $N_{\rm eff}=3.045$, while the case with $\varepsilon_{ee}\neq 0$ gives a smaller value of $N_{\rm eff}$. An explanation of this effect in terms of the effective decoupling temperature of neutrinos can be found in \cite{Mangano:2006ar} (see in particular their Figs.~3 and 4).

\begin{figure}
\centering 
\includegraphics[width=.99\textwidth]{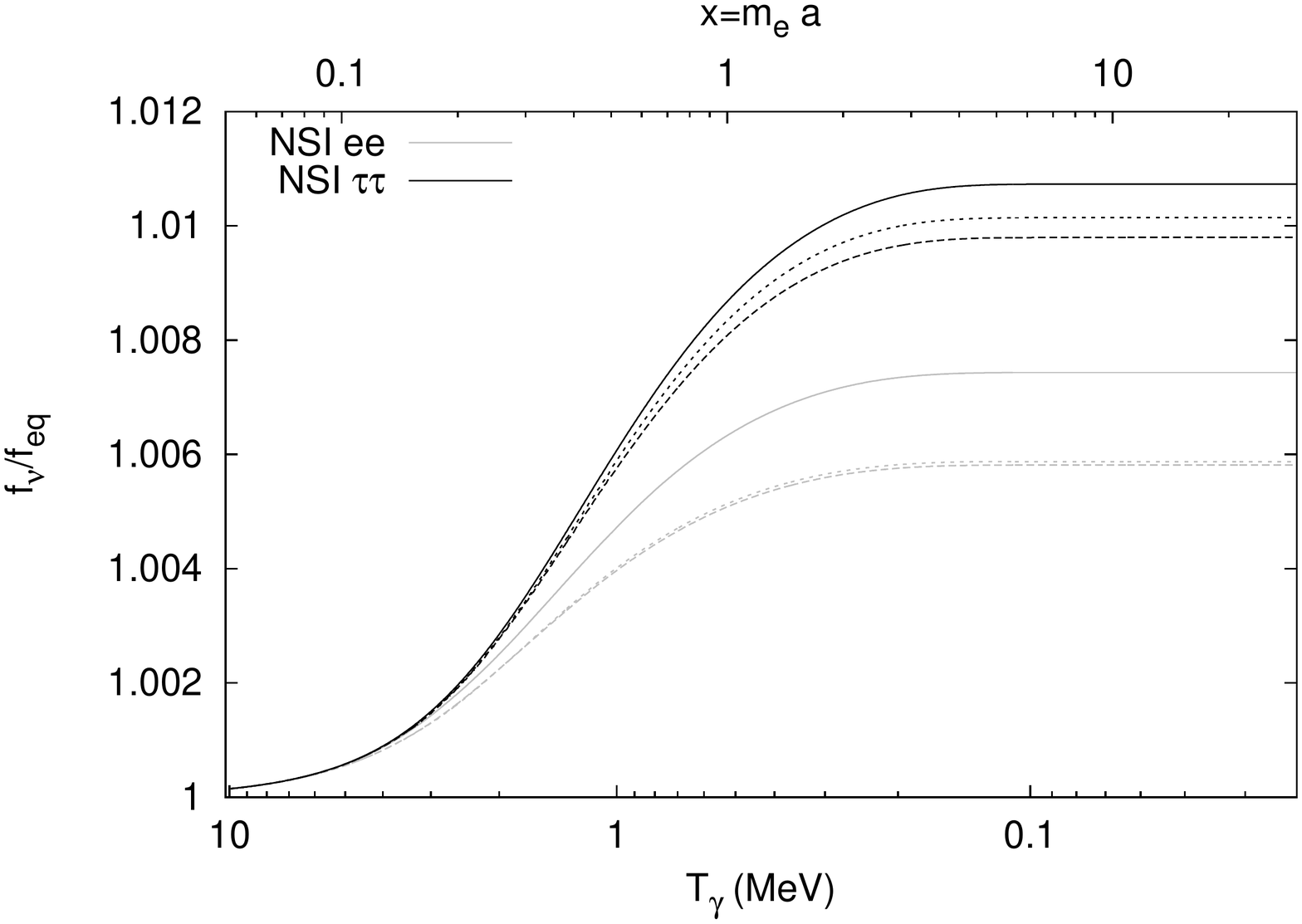}
\caption{\label{fig:evol_nsi} Same as Figure \ref{fig:evol_i} for the two cases with non-standard neutrino interactions, including flavour oscillations and masses in the NH case.}
\end{figure}

\begin{figure}
\centering 
\includegraphics[width=.99\textwidth]{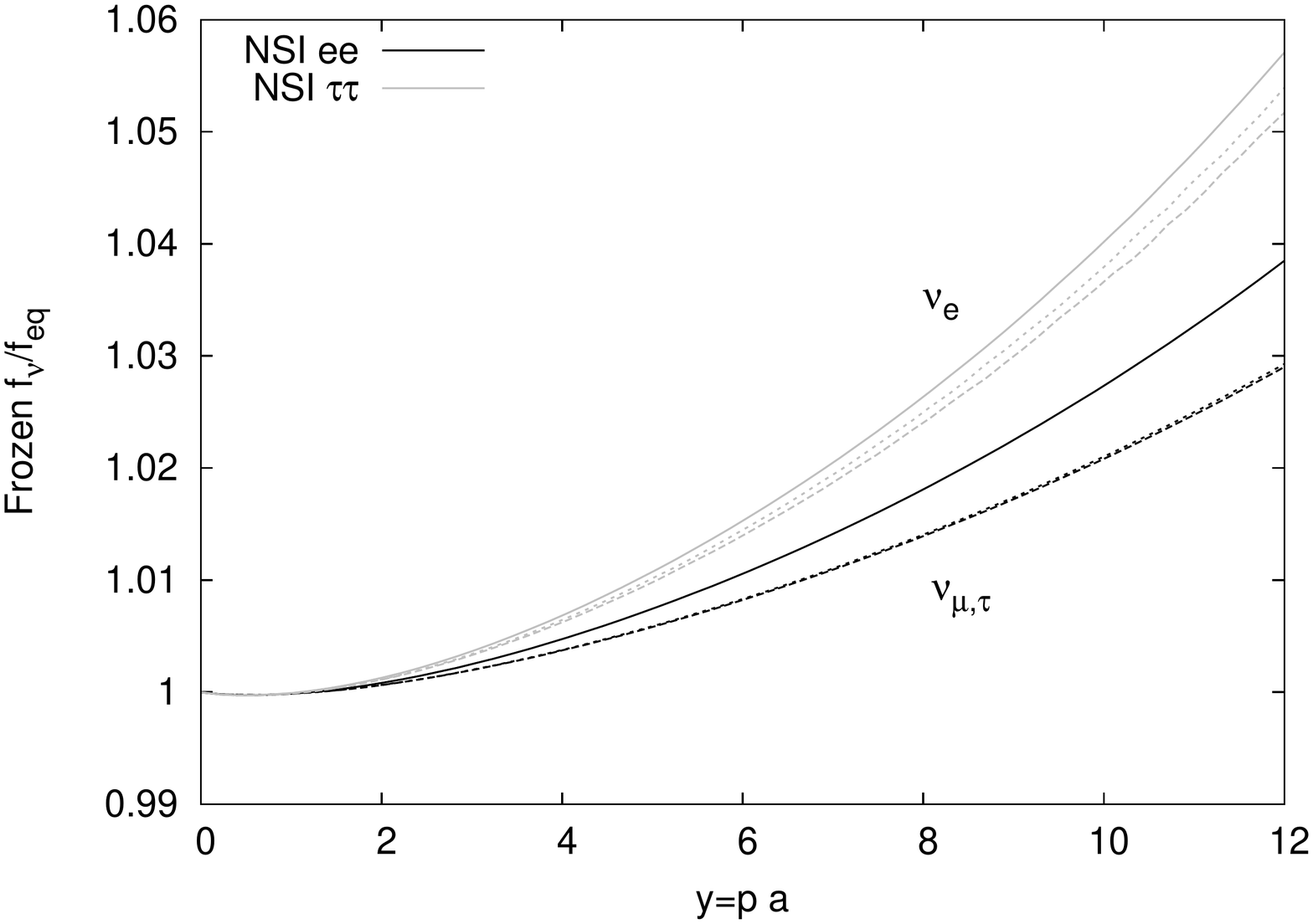}
\caption{\label{fig:delta_nsi} Same as Figure \ref{fig:delta_alpha} for the two cases with non-standard neutrino interactions, including flavour oscillations and masses in the NH case.}
\end{figure}

We show in Fig.\ \ref{fig:evol_nsi} the evolution of the neutrino spectral distortions $f_{\nu_{\alpha}}/f_{\rm eq}$ as a function of the photon temperature or our comoving variable $x$, for both sets of NSI parameters and for a particular neutrino momentum ($y=5$). The asymptotic values of the flavour neutrino distortions as a function of $y$ are depicted in Fig.~\ref{fig:delta_nsi}. The final values of the dimensionless photon temperature, neutrino energy distortions and $N_{\rm eff}$ are listed in the
lower rows of Tables \ref{table_alpha} (flavour neutrino states) and \ref{table_i} (mass neutrino states). The presence of non-zero NSI parameters modifies the neutrino spectral distortions and leads to small changes in the final value of $N_{\rm eff}$: 
$+0.014$ for the case $\varepsilon_{\tau\tau}\neq 0$ and $-0.005$ when $\varepsilon_{ee}\neq 0$.

\begin{table}[tbp]
\centering
\begin{tabular}{|l|ccccc|}
\hline
Case & $z_{\rm fin}$ & $\delta\bar{\rho}_{\nu_1}$(\%) & $\delta\bar{\rho}_{\nu_2}$(\%) & 
$\delta\bar{\rho}_{\nu_3}$(\%) & $N_{\rm eff}$ \\
\hline
Normal hierarchy (NH)     & $1.39779$ & $0.636$ &  $0.574$ & $0.518$ & $3.045$\\
Inverted hierarchy (IH)      & $1.39779$ & $0.635$ & $0.574$ & $0.520$ & $3.045$\\
\hline
NSI $ee$ (NH) & $1.39812$ & $0.548$ & $0.504$ & $0.465$ & $3.040$\\
NSI $\tau\tau$ (NH) & $1.39708$ & $0.839$ & $0.821$ & $0.798$ & $3.059$\\
\hline
\end{tabular}
\caption{\label{table_i} Same as the Tab.~\ref{table_alpha} for the mass neutrino states.}
\end{table}

\section{Conclusions}

In this paper we have revisited the decoupling process of neutrinos in the early universe in the presence of three-flavour oscillations, both with standard weak interactions and with non-standard neutrino interactions with electrons. We have calculated the evolution of the momentum spectra of neutrinos and found their final contribution to the cosmological energy density in the form of radiation, given by the effective number of neutrinos, $N_{\rm eff}$.

This problem requires the numerical solution of the corresponding momentum-dependent kinetic equations for the neutrino density matrix, that we have performed including the full collision terms for neutrino-electron processes. In the case of the off-diagonal components, these collision integrals replace the damping terms usually assumed in previous analyses. We find that, while this improved calculation does change the evolution of these off-diagonal elements, it does not modify the final
value of the effective number of neutrinos, given by $N_{\rm eff}=3.045$. The impact of including the off-diagonal collision terms calculated with the full integrals or the simple damping  prescription could be more important for other cases, such as active-sterile neutrino 
oscillations \cite{Hannestad:2015tea} or very low-reheating scenarios \cite{deSalas:2015glj}, where the spectral flavour distortions could be significantly larger than in the standard case.

We have also considered whether the value of $N_{\rm eff}$ depends on the neutrino mass ordering (normal or inverted), which is still unknown from oscillation experiments. We find very small differences in the final distortions of the flavour and mass neutrino spectra for the two cases, but the same final value of $N_{\rm eff}=3.045$, in agreement with previous theoretical calculations \cite{Mangano:2005cc} and fully consistent with the latest analysis of Planck data \cite{Ade:2015xua}.

In absence of neutrino mixing the process of neutrino decoupling is less complicated. We have also considered this case, currently non-standard, because it can be compared with other previous analyses. Concerning the effective number of neutrinos, we again find $N_{\rm eff}=3.045$, a similar result to the ones obtained in refs.\ \cite{Mangano:2005cc,Birrell:2014uka,Grohs:2015tfy}. It is possible that some of the small differences could arise from the different methods for the numerical evaluation, but we believe that a future specific study of the effects of finite temperature QED radiative corrections on relic neutrino decoupling could be useful. In our analysis we have not calculated the small effects of neutrino heating on primordial nucleosynthesis, and in particular on the production of deuterium or $^4$He, but it has been discussed in refs.\ \cite{dolgov_hansen,esposito,Mangano:2005cc,Grohs:2015tfy}.

Fixing the contribution of neutrinos to the radiation energy density in the standard case that exceeds $N_{\rm eff}=3$ is an important input for the cosmological model. Other relativistic relics or non-standard neutrino properties could lead to a different value of $N_{\rm eff}$, usually larger than 
$3.045$. For instance, in our analysis we have also considered the presence of non-standard neutrino-electron interactions. For two sets of NSI parameters allowed by present data, we find that the final value of $N_{\rm eff}$ can be reduced down to $3.040$ or enhanced up to $3.059$. 

The small departures of $N_{\rm eff}$ from 3 that we have found in this paper, either in the standard case of neutrino decoupling with flavour oscillations or with NSI, are beyond the current level of observational precision fixed by the analysis of Planck data (alone or in combination with other cosmological observations), of the order
of $\sigma(N_{\rm eff})\simeq 0.2$ \cite{Ade:2015xua}. However, forthcoming data from future CMB experiments or large-volume galaxy surveys could improve the sensitivity on $N_{\rm eff}$ down to values such as $0.02-0.04$ (see e.g.\ 
\cite{Basse:2013zua,Abazajian:2013oma,Wu:2014hta}).

\acknowledgments
We thank Omar G.\ Miranda for fruitful discussions on non-standard neutrino interactions. 
P.F.\ de Salas thanks the Institute for Theoretical Particle Physics and Cosmology for hospitality and support when this  paper was being finished. Work supported by the Spanish grants 
FPA2014-58183-P, FPA2015-68783-REDT, Multidark CSD2009-00064 and SEV-2014-0398
(MINECO), FPU13/03729 (MECD) and PROMETEOII/2014/084 (Generalitat Valenciana). 

\bigskip

\appendix

\section{Boltzmann equations and collision terms in comoving variables}\label{Pfunc}

In terms of the comoving variables in eq.\ \eqref{comoving} one can write the Boltzmann equations \eqref{boltzmann} as
\begin{equation}
\frac{d\varrho_y}{dx} =m_{\rm P}\sqrt{\frac{3}{8\pi \bar{\rho}}}\left \{
 -i\frac{x^2}{m^3_e}
\left[ \left( \frac{\mathbb{M}_{\rm F}}{2y} -\frac{8\sqrt{2}G_{\rm F} y m^6_e}{3m^2_{\rm W} x^6}\,\bar{\rho}_{ee} \mathbb{E} \right) ,\varrho_y \right] + \frac{m_e^3}{x^4}\, \bar{\mathcal{I}} \left[ \varrho_y \right]\right \},
\end{equation}
%
where the bar over some quantities indicate that they are written in the comoving variables of eq.\ \eqref{comoving}, such as the dimensionless energy density $\bar{\rho}= {\rho}(x/m_e)^4$. On the other hand, the continuity equation \eqref{EnergyCons} can be translated into an equation for $dz/dx$ as in \cite{Mangano:2001iu}, including the contributions of finite temperature QED corrections.

Here we also present the two-dimensional collision integrals for neutrino-electron processes written in terms of the comoving variables. For the scattering with electrons one has
\begin{align}
\bar{\mathcal{I}}_{\nu e^- \rightarrow \nu e^-} &= \frac{G_{\rm F}^2}{(2\pi)^3 y_1^2} \int dy_2 dy_4 \frac{y_2}{\bar{E}_2} \frac{y_4}{\bar{E}_4} \left\{ \Pi_2 (y_1, y_4) F^{RR}_{\rm sc} (\nu^{(1)}, e^{(2)},\nu^{(3)}, e^{(4)}) \right. \nonumber\\
& + \Pi_2 (y_1,y_2) F^{LL}_{\rm sc}(\nu^{(1)}, e^{(2)},\nu^{(3)}, e^{(4)}) \nonumber\\
& \left. -\, (x^2 +\delta \bar{m}_e^2) \; \Pi_1 (y_1, y_3) \left( F^{RL}_{\rm sc}(\nu^{(1)}, e^{(2)},\nu^{(3)}, e^{(4)})+F^{LR}_{\rm sc}(\nu^{(1)}, e^{(2)},\nu^{(3)}, e^{(4)}) \right) \right\},
\end{align}
where $\bar{E}_i=\sqrt{x^2+\delta \bar{m}_e^2+y_i^2}$ and $\delta \bar{m}_e$ is the finite temperature QED correction to the electron mass (Eq.~(12) in \cite{Mangano:2001iu} expressed in our comoving variables in eq.\ \eqref{comoving}). In the case of the scattering with positrons the matrix with collision terms is 
\begin{align}
\bar{\mathcal{I}}_{\nu e^+ \rightarrow \nu e^+} &= \frac{G_{\rm F}^2}{(2\pi)^3 y_1^2} \int dy_2 dy_4 \frac{y_2}{\bar{E_2}} \frac{y_4}{\bar{E_4}} \left\{ \Pi_2 (y_1, y_4) F^{LL}_{\rm sc} (\nu^{(1)}, \bar{e}^{(2)},\nu^{(3)}, \bar{e}^{(4)}) \right. \nonumber\\
& + \Pi_2 (y_1,y_2) F^{RR}_{\rm sc}(\nu^{(1)}, \bar{e}^{(2)},\nu^{(3)}, \bar{e}^{(4)}) \nonumber\\
& \left. -\, (x^2 +\delta \bar{m}_e^2) \; \Pi_1 (y_1, y_3) \left( F^{RL}_{\rm sc}(\nu^{(1)}, \bar{e}^{(2)},\nu^{(3)}, \bar{e}^{(4)})+F^{LR}_{\rm sc}(\nu^{(1)}, \bar{e}^{(2)},\nu^{(3)}, \bar{e}^{(4)}) \right) \right\},
\end{align}
and for the annihilation to an $e^\pm$ pair,
\begin{align}
\bar{\mathcal{I}}_{\nu \bar{\nu} \rightarrow e^- e^+} &= \frac{G_{\rm F}^2}{(2\pi)^3 y_1^2} \int dy_3 dy_4 \frac{y_3}{\bar{E_3}} \frac{y_4}{\bar{E_4}} \left\{ \Pi_2 (y_1, y_4) F^{LL}_{\rm ann} (\nu^{(1)}, \bar{\nu}^{(2)},e^{(3)}, \bar{e}^{(4)}) \right. \nonumber\\
& + \Pi_2 (y_1,y_3) F^{RR}_{\rm ann} (\nu^{(1)}, \bar{\nu}^{(2)},e^{(3)}, \bar{e}^{(4)}) \nonumber\\
& \left. +\, (x^2 +\delta \bar{m}_e^2) \; \Pi_1 (y_1,y_2) \left( F^{RL}_{\rm ann} (\nu^{(1)}, \bar{\nu}^{(2)},e^{(3)}, \bar{e}^{(4)}) + F^{LR}_{\rm ann} (\nu^{(1)}, \bar{\nu}^{(2)},e^{(3)}, \bar{e}^{(4)}) \right) \right\}.
\end{align}
%
Finally, the functions $\Pi_{1,2} (y_i,y_j)$ have the following structure:
\begin{equation}
\Pi_1(y_1,y_3) = \bar{E}_1 \bar{E}_3 D_1 + D_2(y_1,y_3,y_2,y_4)
\end{equation}
\begin{equation}
\Pi_1(y_1,y_2) = \bar{E}_1 \bar{E}_2 D_1 - D_2(y_1,y_2,y_3,y_4)
\end{equation}
\begin{equation}
\Pi_2(y_1,y_4) = 2\left( \bar{E}_1 \bar{E}_2 \bar{E}_3\bar{E}_4 D_1 + \bar{E}_2 \bar{E}_3 D_2(y_1,y_4,y_2,y_3) + \bar{E}_1 \bar{E}_4 D_2 (y_2,y_3,y_1,y_4) + D_3  \right)
\end{equation}
\begin{equation}
\Pi_2(y_1,y_2) = 2\left( \bar{E}_1 \bar{E}_2 \bar{E}_3\bar{E}_4 D_1 - \bar{E}_1 \bar{E}_2 D_2(y_3,y_4,y_1,y_2) - \bar{E}_3 \bar{E}_4 D_2 (y_1,y_2,y_3,y_4) + D_3  \right)
\end{equation}
\begin{equation}
\Pi_2(y_1,y_3) = 2\left( \bar{E}_1 \bar{E}_2 \bar{E}_3\bar{E}_4 D_1 + \bar{E}_1 \bar{E}_3 D_2(y_2,y_4,y_1,y_3) + \bar{E}_2 \bar{E}_4 D_2 (y_1,y_3,y_2,y_4) + D_3  \right)
\end{equation}
where
\begin{equation}
\bar{E_i}=\begin{cases}
\sqrt{x^2+\delta \bar{m}_e^2+y_i^2}\qquad\quad & \mathrm{for\; electrons\; or\; positrons},\\
\qquad y_i \qquad\qquad\qquad & \mathrm{for\; neutrinos}.
\end{cases}
\end{equation}
The $D_l$ functions are obtained as in \cite{dolgov_hansen} from the following integrals:
\begin{equation}
D_1= \frac{16}{\pi}\int_0^\infty \frac{d\lambda}{\lambda^2} \sin(\lambda y_1) \sin (\lambda y_2) \sin(\lambda y_3) \sin(\lambda y_4)
\end{equation}
\begin{eqnarray}
D_2(y_i,y_j,y_k,y_l) &=& - \frac{16}{\pi} \int_0^\infty \frac{d\lambda}{\lambda^4} \sin(\lambda y_k)\sin (\lambda y_l) \left[ \lambda y_i \cos(\lambda y_i) - \sin (\lambda y_i) \right]\nonumber \\
&&\times \left[ \lambda y_j \cos(\lambda y_j) - \sin (\lambda y_j) \right] 
\end{eqnarray}
\begin{equation}
D_3 = \frac{16}{\pi} \int_0^\infty \frac{d\lambda}{\lambda^6} \prod_{a=1}^4 \left[ \lambda y_a \cos(\lambda y_a) - \sin (\lambda y_a) \right]  
\end{equation}
that can be solved analytically. We have not explicitly written the dependence on $y_i$ of $D_1$ and $D_3$ because they are symmetric in the dimensionless neutrino momenta. 

\bibliography{references1}
\bibliographystyle{JHEP}

\end{document}